\documentclass[superscriptaddress,twocolumn]{revtex4}
\usepackage{mathrsfs}
\usepackage{amssymb}
\usepackage[tbtags]{amsmath}
\usepackage{graphicx}
\usepackage{epsfig,graphicx,times}
\usepackage{color}
\usepackage{subfigure}

\begin{document}

\title{Coherently manipulating cold ions in separated traps
by their vibrational couplings}
\author{Miao Zhang}
\affiliation{Quantum Optoelectronics Laboratory, School of Physics
and Technology, Southwest Jiaotong University, Chengdu 610031,
China}
\author{L. F. Wei\footnote{weilianfu@gmail.com}}
\affiliation{Quantum Optoelectronics Laboratory, School of Physics
and Technology, Southwest Jiaotong University, Chengdu 610031,
China} \affiliation{State Key Laboratory of Optoelectronic Materials
and Technologies, School of Physics and Engineering, Sun Yat-sen
University, Guangzhou 510275, China} \affiliation{State Key
Laboratory of Functional Materials for Informatics, Shanghai
Institute of Microsystem and Information Technology, Chinese Academy
of Sciences, Shanghai 200050 China}
\date{\today}

\begin{abstract}
Recent experiments [K. R. Brown, {\it et al}., Nature  471, 196
(2011); and M. Harlander, {\it et al}., Nature 471, 200 (2011)] have
demonstrated the coherent manipulations on the external vibrations
of two ions, confined individually in the separated traps. Using
these recently developed techniques, we propose here an approach to
realize the coherent operations, e.g., the universal quantum gates,
between the separated ion-trap qubits encoded by two internal atomic
states of the ions. Our proposal operates beyond the usual
Lamb-Dicke limits, and could be applied to the scalable ion-traps
coupled by their vibrations.

\end{abstract}

\maketitle

In the past two decades, much attention has been paid to the study
of quantum computer, which uses the principles of quantum mechanics
to solve problems that could never be solved by any classical
computer~\cite{Shor}. Up to now, several kinds of physical systems,
e.g., trapped ions~\cite{IONS}, cavity QEDs~\cite{QED}, nuclear
magnetic resonance~\cite{NMR}, Josephson junctions~\cite{JJ}, and
coupled quantum dots~\cite{CQD}, etc., have been proposed to
implement the desirable quantum computation. Specifically, the
system of trapped ions has many advantages such as convenient
manipulation, relatively long coherence time, and easy
readout~\cite{IONS2}.

Quantum computation with trapped ions was proposed by Cirac and
Zoller~\cite{IONS}. In their scheme, a string of ions is trapped in
{\it a single potential well}, and the ions' collective motions
(CMs) act as the data bus to couple the distant qubits encoded by
the internal atomic states of ions. The traditional Cirac-Zoller
gate~\cite{IONS} requires that the CMs of trapped ions should be
cooled to their vibrational ground state; thus the implementation of
such a gate is practically sensitive to the decoherence of the
motional states. Subsequently, Milburn~\cite{warm-ion1} and
M${\text{\o}}$lmer and S${\text{\o}}$rensen~\cite{warm-ion2}
proposed alternative models of quantum computation with warm ions,
which relax the stringent requirements on cooling vacuum state and
addressing individual ion.
Recently, there has been much interest in {\it multi-zone traps} for
the implementation of a scalable quantum computing
network~\cite{Cirac2000,scalable}, in which ions are confined
individually in separated traps and coupled by their Coulomb
interactions. Recently this idea has been demonstrated
experimentally~\cite{experiment1,experiment2} by directly
controlling trapping potentials (i.e., the voltages on the DC
electrodes) to tune their external vibrations for resonances.

Following the above experimental
demonstrations~\cite{experiment1,experiment2}, in this brief report
we propose an approach to implement a typical coherent manipulation,
i.e., a universal controlled-NOT (CNOT) gate, between the separated
ion traps. It is well known that such a universal gate, assisted by
arbitrary single-qubit rotations, can generate any quantum computing
network~\cite{CNOT}. In our approach, several laser pulses are
applied to the trapped ions to exchange the information between the
external and internal states of the ions. The switchable coupling
between the separately trapped ions is achieved by adiabatically
sweeping the trapping potentials. Here, the adiabaticity means that
such a sweeping does not yield any quantum transition between the
ions' vibrational levels. Our proposal operates beyond the usual
Lamb-Dicke (LD) limit~\cite{LFWei,OC} and could be scaled to more
than two qubits.

First, we simply review how to realize the laser-induced coupling
between the external and internal states of a single trapped ion. A
single trapped ion (with mass $M$ and charge $q$) has two degrees of
freedom: the vibrational motion around the trap center and the
internal atomic levels. We assume that the ion trap provides a
pseudopotential such that the ion's oscillation frequency $\nu$
along the axial direction is much smaller than those along the
radial directions~\cite{experimentstate,experimentgate}. As a
consequence, only the quantized vibrational motion along the axial
direction is considered.
For the ion's internal degrees of freedom we consider two atomic
levels, e.g., the ground state $|g\rangle$ and the excited state
$|e\rangle$, to encode a qubit. The Hamiltonian describing the two
degrees of freedom of the ion reads
\begin{equation}
\hat{H}_0=\hbar\nu(\hat{a}^\dagger\hat{a}+\frac{1}{2})+\frac{\hbar}{2}\omega_a\hat{\sigma_z},
\end{equation}
where $\hat{a}^\dagger$ and $\hat{a}$ are the bosonic creation and
annihilation operators of the external vibrations and
$\hat{\sigma}_z=|e\rangle\langle e|-|g\rangle\langle g|$ is the
Pauli operator of the qubit. The transition frequency $\omega_a$ is
defined by $\omega_a=(E_e-E_g)/\hbar$, with $E_g$ and $E_e$ being
the corresponding energies of the ground and excited states,
respectively.
It is well known that the coupling between the above two uncoupled
degrees of freedom of the ion can be achieved by applying suitable
classical lasers. This coupling can be descried by the effective
laser-ion interaction~\cite{LFWei,OC}
\begin{equation}
\hat{H}_i=\frac{\hbar\Omega}{2}(\hat{\sigma}_++\hat{\sigma}_-)
e^{i\eta(\hat{a}+\hat{a}^\dagger)-i\omega_lt-i\vartheta_l}+\text{H.c.},
\end{equation}
where $\Omega$ is the Rabi frequency describing the strength of the
coupling between the applied lasers and the trapped ion, $\eta$ is
the LD parameter describing the strength of the coupling between the
external and internal degrees of freedom of the ion, $\omega_l$ and
$\vartheta_l$ are the effective frequency and initial phase of the
applied laser beams, respectively, and
$\hat{\sigma}_+=|e\rangle\langle g|$ and
$\hat{\sigma}_-=|g\rangle\langle e|$ are the usual raising and
lowering operators, respectively.

If the LD parameter $\eta$ is sufficiently small, then the usual LD
approximation works well and thus, by neglecting the terms relating
the higher order of $\eta$,
$\exp\left[i\eta(\hat{a}+\hat{a}^\dagger)\right]
\approx1+i\eta(\hat{a}+\hat{a}^\dagger)$~\cite{experimentstate,experimentgate}.
Beyond such a limit, the total Hamiltonian
$\hat{H}=\hat{H}_0+\hat{H}_i$ of the trapped ion can be generally
written as
\begin{equation}
\begin{array}{l}
\hat{H}'=\frac{\hbar\Omega}{2}e^{-\eta^2/2}
\hat{\sigma}_+\sum_{n,m}^\infty\left[(-1)^{n+m}f_1+ f_2\right]
\\
\\
\,\,\,\,\,\,\,\,\,\,\,\,\,\,\,\times\frac{(i\eta)^{n+m}
(\hat{a}^\dagger)^n\hat{a}^m e^{i(n-m)\nu t}}{n!m!}+\text{H.c.}
\end{array}
\end{equation}
in the interaction picture defined by the unity operator
$\hat{U}_1=\exp(-it\hat{H}_0/\hbar)$, where
$f_1=\exp[i(\omega_a+\omega_l)t+i\vartheta_l]$ and
$f_2=\exp[i(\omega_a-\omega_l)t-i\vartheta_l]$.
Specifically, for the $k$th red-sideband excitations, i.e.,
$\omega_l=\omega_a-k\nu$ with $k=0,1,2,\cdot\cdot\cdot$, the above
Hamiltonian reduces effectively to
\begin{equation}
\hat{H}_r=\frac{\hbar\Omega}{2}e^{-\eta^2/2-i\vartheta_l}
(i\eta)^k\hat{\sigma}_+ \sum_{j=0}^\infty
\frac{(i\eta)^{2j}(\hat{a}^\dagger)^j\hat{a}^{j+k}}{j!(j+k)!}
+\text{H.c.}
\end{equation}
in the rotating-wave approximation [i.e., neglecting the rapid
oscillating terms in (3)].
The dynamics of the trapped ion ruled by Hamiltonian (4) is exactly
solvable. Indeed, if the external vibration of the ion is initially
in a Fock state $|m\rangle$ and the internal atomic state is in the
ground sate $|g\rangle$ (or excited state $|e\rangle$), then the
relevant state evolutions can be given by
\begin{eqnarray}
\left\{
\begin{array}{l}
|m,g\rangle\longrightarrow|m,g\rangle,\,\,m<k,\\
\\
|m,g\rangle\longrightarrow\cos(\Omega_{m-k,k}t)|m,g\rangle
\\
\,\,\,\,\,\,\,\,\,\,\,\,\,\,\,\,\,\,\,\,\,\,\,\,\,\,
+i^{k-1}e^{-i\vartheta_l}\sin(\Omega_{m-k,k}t)
|m-k,e\rangle,\,\,m\geq
k,\\
\\
|m,e\rangle\longrightarrow\cos(\Omega_{m,k}t)|m,e\rangle
\\
\,\,\,\,\,\,\,\,\,\,\,\,\,\,\,\,\,\,\,\,\,\,\,\,\,\,
-(-i)^{k-1}e^{i\vartheta_l}\sin(\Omega_{m,k}t)|m+k,g\rangle,
\end{array}
\right.
\end{eqnarray}
with
\begin{equation}
\Omega_{m,k}=\frac{\Omega}{2}\eta^ke^{-\eta^2/2}\sqrt{\frac{(m+k)!}{m!}}\sum_{j=0}^m\frac{(i\eta)^{2j}m!}{(j+k)!j!(m-j)!}\,.
\end{equation}
\begin{figure}[tbp]
\includegraphics[width=8cm]{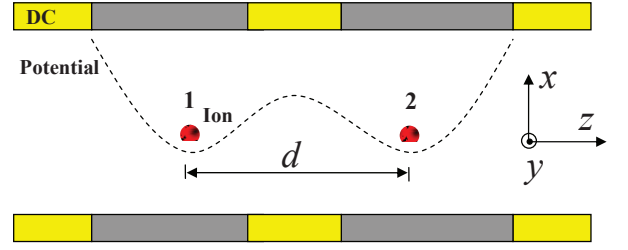}
\caption{(Color online) Sketch of two ions confined individually in
two potential wells  separated by a distance $d$ between them.}
\end{figure}

We now consider two ions are trapped individually in two potential
wells  separated by a distance $d$ between
them~\cite{experiment1,experiment2}, as shown in Fig.1. The
Hamiltonian describing the external motions of the two ions reads
\begin{equation}
\hat{H}_{ex}=\hbar\nu_1(\hat{a}_1^\dagger\hat{a}_1+\frac{1}{2})
+\hbar\nu_2(\hat{a}_2^\dagger\hat{a}_2+\frac{1}{2})+V,
\end{equation}
with
\begin{equation}
V\approx -\frac{q_1q_2}{4\pi\epsilon_0d}\left[1+\frac{z_1-z_2}{d}
-\frac{(z_1-z_2)^2}{d^2}\right]
\end{equation}
being the Coulomb interaction between them. Here, $z_1$ and $z_2$
are the displacements of the trapped ions from the minimal values of
the two potentials, respectively. In terms of the bosonic operators
introduced above, the Coulomb interaction between the two ions can
be rewritten as
\begin{equation}
\begin{array}{l}
V=\mathcal{K}\bigg[\sum_{j=1}^2\frac{(-1)^{j}\xi_j}{d}\hat{a}_j
+\frac{\xi_j^2}{d^2}(\hat{a}_j^2+\hat{a}_j\hat{a}_j^{\dagger})
\\
\\
\,\,\,\,\,\,\,\,\,\,\,\,\, -\frac{2\xi_1\xi_2}{d^2}
(\hat{a}_1\hat{a}_2+\hat{a}_1\hat{a}_2^\dagger)\bigg] +\text{H.c.},
\end{array}
\end{equation}
with $\mathcal{K}=q_1q_2/(4\pi\epsilon_0d)$ and
$\xi_j=\sqrt{\hbar/(2M_j\nu_j)}$. Consequently, the Hamiltonian (7)
can be further written as
\begin{equation}
\begin{array}{l}
\hat{H}'_{ex}=\mathcal{K}\bigg[
\sum_{j=1}^2\frac{(-1)^{j}\xi_j}{d}\hat{a}_je^{-i\omega_jt}
+\frac{\xi_j^2}{d^2}\hat{a}_j^2e^{-i2\omega_jt}
\\
\\
\,\,\,\,\,\,\,\,\,\,\,\,\,\,\,\,\,\,-\frac{2\xi_1\xi_2}{d^2}
(\hat{a}_1\hat{a}_2e^{-i(\omega_1+\omega_2)t}
+\hat{a}_1\hat{a}_2^\dagger e^{i(\omega_2-\omega_1)t})\bigg]
+\text{H.c.}
\end{array}
\end{equation}
in the interaction picture defined by unity operator
$\hat{U}_2=\exp(-it\hat{H}^0_{ex}/\hbar)$. Here,
$\hat{H}^0_{ex}=\sum_{j=1}^2\hbar\omega_j(\hat{a}_j^\dagger
\hat{a}_j+1/2)$, and $\omega_j=\nu_j+2\mathcal{K}\xi_j^2/(\hbar
d^2)$ is the renormalized frequency of the $j$th ion.
Usually, the amending frequency
$\tilde{\nu}_j=2\mathcal{K}\xi_j^2/(\hbar d^2)$ is much smaller than
the original frequency $\nu_j$ and thus could be negligible. For
example, for the experimental parameters:
$\nu_j=2\pi\times4.04$~MHz, $d=40\,\mu$m, and mass
$M_j=M(^9\text{Be}^+)$~\cite{experiment1}, we have
$\tilde{\nu}_j\approx2\pi\times1.5$~KHz.

Under the resonant condition, i.e., $\omega_1=\omega_2$, the
Hamiltonian (10) can be reduced effectively to
\begin{equation}
\hat{H}_{\text{eff}}=-\hbar
g(\hat{a}_1\hat{a}_2^\dagger+\hat{a}_1^\dagger\hat{a}_2)
\end{equation}
under the rotating wave approximation. Where, the coupling strength
$g=2\mathcal{K}\xi_1\xi_2/d^2$. Consequently, in the lowest two
invariant subspaces $\{|0\rangle_1|0\rangle_2\}$ and
$\{|0\rangle_1|1\rangle_2\rangle, |1\rangle_1|0\rangle_2\}$, the
dynamics of the coupled bosonic modes reads
\begin{equation}
|0\rangle_1|0\rangle_2\longrightarrow|0\rangle_1|0\rangle_2,
\end{equation}
and
\begin{eqnarray}
\left\{
\begin{array}{l}
|0\rangle_1|1\rangle_2\longrightarrow\cos(gt)|0\rangle_1|1\rangle_2
+i\sin(gt)|1\rangle_1|0\rangle_2,\\
\\
|1\rangle_1|0\rangle_2\longrightarrow\cos(gt)|1\rangle_1|0\rangle_2
+i\sin(gt)|0\rangle_1|1\rangle_2.
\end{array}
\right.
\end{eqnarray}

Suppose that the two ions are initially trapped in the separated
potential wells with different frequencies (i.e., without
resonance), and their external vibrational states are both prepared
in the vacuum state $|0\rangle$. Our approach, to realize the
desirable CNOT gate between the internal atomic states of two ions,
includes the following five sequential operations.

(I) A first red-sideband pulse with the duration
$t_1=\pi/(2\Omega_{0,1})$ and phase $\vartheta_1$ is applied on the
first ion to implement the operation $|0,e\rangle_1\longrightarrow
\exp[i(\vartheta_1+\pi)]|1,g\rangle_1$, leaving the state
$|0,g\rangle_1$ unchanged.

(II) Varying adiabatically the trapping potentials such that the two
vibrations are turned into resonance. After a duration
$t_2=\pi/(2g)$, this resonant coupling is then adiabatically turned
off and thus the evolution $|1\rangle_1|0\rangle_2\longrightarrow
e^{i\pi/2}|0\rangle_1|1\rangle_2$ is  generated, leaving the state
$|0\rangle_1|0\rangle_2$ unchanged.

(III) A resonant pulse with the duration $t_3$ and phase
$\vartheta_3$ is applied on the second ion to implement the
single-ion CNOT gate between the external and internal states of the
ion~\cite{OC}. This is the key step for generating the desirable
CNOT gate between the two separately trapped ions. The
implementation of such a single-ion CNOT gate (beyond the usual LD
regime) by using a single resonant pulse will be further explained
later in the text.

(IV) Repeating the operations in step (II).

(V) Another first red-sideband pulse with the duration
$t_5=\pi/(2\Omega_{0,1})$ and phase $\vartheta_5$ is applied to the
first ion.

After the above five sequential manipulations, we have
\begin{equation}
\begin{array}{l}
|0,g\rangle_1|0,g\rangle_2
\longrightarrow|0,g\rangle_1|0,g\rangle_2\\
|0,g\rangle_1|0,e\rangle_2
\longrightarrow|0,g\rangle_1|0,e\rangle_2\\
|0,e\rangle_1|0,g\rangle_2\longrightarrow
e^{i(\vartheta_1-\vartheta_3-\vartheta_5+3\pi/2)}
|0,e\rangle_1|0,e\rangle_2\\
|0,e\rangle_1|0,e\rangle_2 \longrightarrow e^{i(\vartheta_1
+\vartheta_3-\vartheta_5+3\pi/2)}|0,e\rangle_1|0,g\rangle_2.
\end{array}
\end{equation}
Certainly, if the phases of the applied laser beams satisfy the
conditions $\vartheta_1-\vartheta_3-\vartheta_5=\vartheta_1
+\vartheta_3-\vartheta_5=-3\pi/2$, then the qubits undergo a
desirable CNOT logic gate and their vibrations return to the ground
states.

We now discuss the conditions of the adiabatic manipulations in
steps II and IV for switching on/off the ion-ion couplings. To
realize these adiabatic operations, an possible method is to vary
the vibrational frequency of one of the two ions, e.g., the first
one, as $\nu_{1}(\tau)=\beta\tau+\nu_2$ with the rate $\beta$ and
time $\tau$. This indicates that the detuning
$\Delta=\nu_{1}(\tau)-\nu_2 =\beta\tau$ between the two vibrations
of the ions are controllable (although the frequency $\nu_2$ of the
second ion is fixed). As a consequence, the coupling between the two
ions could be turned on (with $\Delta=0$) or off (with $\Delta\gg
g$). Obviously, with the time-dependent frequency $\nu_{1}(\tau)$
the Hamiltonian (7) can be rewritten as
\begin{equation}
\hat{H}_{ex}(\tau)=-\frac{\hbar^2}{2m_1}\frac{\partial^2}{\partial
z_1^2}+\frac{1}{2}M_1\nu^2_{1}(\tau)z_1^2
+\hbar\nu_2(\hat{a}_2^\dagger\hat{a}_2+\frac{1}{2})+V.
\end{equation}
According to the standard quantum adiabatic theorem, the occupations
of the vibrational states of the ions are unchanged during the
adiabatic manipulations. For the present case, the condition
\begin{equation}
\gamma_{n,m}=\left|\frac{_1\langle
n|\frac{\partial\hat{H}_{ex}(\tau)}{\partial
\tau}|m\rangle_1}{\hbar\nu^2_{1}(\tau)(n-m)^2}\right|
\ll1,\,\,\,\,\,n\neq m,
\end{equation}
should be satisfied. Indeed, by proper setting the relevant
parameters this condition can be well satisfied, e.g.,
$\gamma_{n,0}<3.1\times10^{-6}$ and $\gamma_{n,1}<5.3\times10^{-6}$
for the typical $\Delta=100$~KHz, $\tau=9\,\mu$s, and
$\nu_{1}(\tau)=2\pi\times4.04$~MHz~\cite{experiment1}.
Consider that two $^9\text{Be}^+$ ions are trapped separately at the
distance $d\approx40\,\mu$m, and the resonant frequency is set at
$\nu_1=\nu_2=2\pi\times4.04$~MHz~\cite{experiment1}. Then, the
coupling strength is caculated as $g\approx2\pi\times1.5$~KHz, which
is far smaller than the initial detuning $\Delta=100$~KHz set above.
Obviously, the selected duration $\tau=9\,\mu$s~\cite{experiment1}
of adiabatic operation is significantly shorter than the periods
$2\pi/g\approx0.67$~ms of the resonant coupling. This indicates that
the influence of the adiabatic manipulation on the ion-ion coupling
is practically weak. As a consequence, the influence on the designed
$\pi$-pulse $\sin(\pi/2+2g\tau)\approx0.99$, applied in steps II and
IV, is practically weak, due to the relatively-fast adiabatic
manipulations. Therefore, the present adiabatic manipulation is slow
enough to satisfy the required adiabatic condition, but is
practically sufficiently fast to realize the ion-ion switchable
couplings.

The single-ion CNOT gate in step III is usually implemented within
the LD limit, see, e.g., Ref~\cite{experimentgate}. In that
experiment, three sequential laser pulses and an auxiliary atomic
level are required. Beyond the LD limit and without the auxiliary
level, we show in the step III that only a single laser pulse is
sufficient to implement the desirable single-ion CNOT gate. Indeed,
under the resonant driving, if the duration $t_3$ of the single
laser pulse is set properly, e.g.,
$\cos(\Omega_{0,0}t_3)=\sin(\Omega_{1,0}t_3)=1$, then the desirable
CNOT gate of a single trapped ion could be implemented. Of course,
the durations $t_3$ are now not that of the usual
$\pi$-pulses~\cite{experimentgate} but other ones, which is resolved
from $\cos(\Omega_{0,0}t_3)=\sin(\Omega_{1,0}t_3)=1$ by the
numerical method~\cite{OC}. Using the Raman beams of
experiment~\cite{experimentstate}, the LD parameter
$\eta\approx0.33$ for the vibrational frequency
$\nu=2\pi\times4.04$~MHz, and $t_3\approx29.6\,\mu$s is resolved for
the Rabi frequency $\Omega=2\pi\times500$~KHz.

For the feasibility, the total durations $t_{\text{total}}$ of the
above five-step operations must be sufficiently short, such that the
coherence of quantum states could not be disturbed significantly by
the practically-existing decoherence. For the typical parameters
listed above: $\eta=0.33$, $\Omega=2\pi\times500$~KHz, and
$g=2\pi\times1.5$~KHz, we have
$t_1=t_5=\pi/(2\Omega_{0,1})\approx3.2\,\mu$s and
$t_2=\pi/(2g)\approx166.7\,\mu$s.
Consequently, the total duration
$t_{\text{total}}=2(t_1+t_2)+t_3+4\tau\approx405.4\,\mu$s. This
implies that the present approach to generate the desirable CNOT
gate should be feasible, since the coherence of the vibrational
states between the two trapped ions could be maintained on the order
of milliseconds~\cite{experiment1}.
In principle, the above durations of operations can be significantly
shortened by increasing the Rabi frequency $\Omega$ and the coupling
strength $g$. Experimentally, these parameters can be enlarged by
increasing the power $P$ of the applied laser beams and decreasing
the distance $d$ between the potential wells (since
$\Omega\propto\sqrt{P}$ and $g\propto1/d^3$). For example, if $P$ is
enhanced to $5$~mW (ten times than that used in
experiment~\cite{experimentstate}) and $d$ is shortened to
$20\,\mu$m (half of that used in~\cite{experiment1}), then the Rabi
frequency $\Omega\approx2\pi\times1.6$~MHz and the coupling strength
$g\approx2\pi\times12$~KHz can be reached. Thus, the total duration
is shortened to $t_{\text{total}}\approx88.9\,\mu$s.

\begin{figure}[tbp]
\includegraphics[width=8cm]{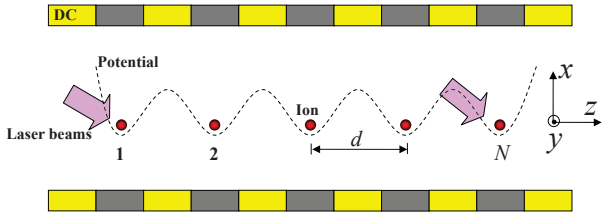}
\caption{(Color online) Sketch of coherent manipulations between two
ions separated by the distance $(N-1)d$.}
\end{figure}

Finally, we discuss the scalability of the system, i.e., how many
ions ($N>2$) confined individually in different traps can be
integrated as a trap-chain (see Fig.~2). For simplicity, we assume
that the distances between the nearest-neighbor ions are uniform
$d$, and all the ions are initially decoupled and cooled in their
motional ground states $\prod_{j=1}^N|0_j\rangle$.
As the distant coupling $g\propto1/[(N-1)d]^3$ (between the first
and $N$th ions) decreases quickly with the increase of the integer
number $N$, the quantum logic operation between them is not easy to
be implemented by using such a direct coupling. A possibly feasible
method is to sequentially implement the operation of resonant
coupling between the nearest-neighbor ions one by one, i.e.,
\begin{equation}
\begin{array}{l}
|1\rangle_1|0\rangle_2\cdots|0\rangle_{N}|Q_N\rangle\longrightarrow
e^{i\pi}|0\rangle_1|1\rangle_2\cdots|0\rangle_{N}|Q_N\rangle
\longrightarrow\cdots
\\
\\
\,\,\,\,\,\,\,\,\,\,\,\,\,\,
\,\,\,\,\,\,\,\,\,\,\,\,\,\,\longrightarrow
e^{i(N-1)\pi}|0\rangle_1|0\rangle_{2}\cdots|1\rangle_{N}|Q_N\rangle.
\end{array}
\end{equation}
By this way the vibrational information of the first ion could be
transferred to the last one without affecting their internal states
(i.e., the $N$-qubit state $|Q_N\rangle$ is unchanged). This means
that the step II should include $2(N-1)$ adiabatic operations for
sequentially coupling the trapped ions one by one. Consequently, the
total duration for generating the CNOT gate between the first and
$N$th ions is $t'_{\text{total}}=
2\left[t_1+(N-1)t_2\right]+t_3+4(N-1)\tau\approx709.7\,\mu$s with
$\Omega=2\pi\times1.6$~MHz, $d=20\,\mu$m, and $N=10$. Based on these
estimations, a few traps (e.g., up to $N\sim 10$) could be
integrated to realize a scaled quantum computing network, at least
theoretically.

In conclusion, we have proposed an approach to realize the universal
CNOT gate between the separately trapped ions. This approach
operates beyond the usual LD limit and includes three steps laser
operations and two steps adiabatic manipulations. The laser beams
are used to couple the external and internal states of trapped ions,
and the adiabatic manipulations are utilized to realize the
switchable coupling between the vibrations of the ions. A possible
method to implement the scalable ion-traps quantum computing
networks is also presented.

{\bf Acknowledgements}: This work is partly supported by the NSFC
grant No. 10874142, 90921010, the grant from the Major State Basic
Research Development Program of China (973 Program) (No.
2010CB923104), and the open project of state key laboratory of
functional materials for informatics.



\begin{thebibliography}{99}

%
\bibitem{Shor}P. W. Shor, in {\it Proceedings of the 35th Annual Symposium on
Foundations of Computer Science}, edited by S. Goldwasser (IEEE
Computer Society Press, Los Alamitos, CA, 1994), p. 124; A. Ekert
and R. Jozsa, Rev. Mod. Phys. {\bf 68}, 733 (1996).
%
\bibitem{IONS}J. I. Cirac and P. Zoller, Phys. Rev. Lett. {\bf 74}, 4091 (1995).
%
\bibitem{QED}J. M. Raimond, M. Brune, and S. Haroche, Rev. Mod. Phys. {\bf 73},
565 (2001); J. I. Cirac, P. Zoller, H. J. Kimble, and H. Mabuchi,
Phys. Rev. Lett. {\bf 78}, 3221 (1997); M. Keller {\it et al.},
Nature {\bf 431}, 1075 (2004).
%
\bibitem{NMR}N. A. Gershenfeld and I. L. Chuang, Science {\bf 275}, 350 (1997).
%
\bibitem{JJ}Y. Makhlin, G. Sch$\text{\"{o}}$n, and A. Shnirman, Rev. Mod. Phys. {\bf 73},
357 (2001).
%
\bibitem{CQD}D. Loss and D. P. DiVincenzo, Phys. Rev. A {\bf 57}, 120 (1998).
%
\bibitem{IONS2}D. Leibfried, R. Blatt, C. Monroe, and
D. Wineland, Rev. Mod. Phys. {\bf 75} 281 (2003); H.
H$\ddot{\text{a}}$ffner {\it et al.}, Nature {\bf 438}, 643 (2005).
%
\bibitem{warm-ion1}G.
J. Milburn, S. Schneider, and D. F. V. James, Fortschr. Phys. {\bf
48} 801 (2000).
%
\bibitem{warm-ion2}K.
M${\text{\o}}$lmer and A. S${\text{\o}}$rensen, Phys. Rev. Lett.
{\bf 82} 1835 (1999).
%
\bibitem{Cirac2000}J. I. Cirac and P. Zoller, Nature {\bf 404}, 579 (2000).
%
\bibitem{scalable}D. Kielpinski, C. Monroe, and
D. J. Wineland, Nature {\bf 417}, 709 (2002).
%
\bibitem{experiment1}K. R. Brown {\it et al.}, Nature {\bf 471},
196 (2011).
%
\bibitem{experiment2}M. Harlander {\it et al.}, Nature {\bf 471}, 200 (2011).
%
\bibitem{CNOT}T. Sleator and H. Weinfurter,
Phys. Rev. Lett. {\bf 74}, 4087 (1995); D. P. DiVincenzo, Phys. Rev.
A {\bf 51}, 1015 (1995).
%
\bibitem{LFWei}
L. F. Wei, S. Y. Liu, and X. L. Lei, Phys. Rev. A, {\bf 65} 062316
(2002); L. F. Wei, Y. X. Liu, and F. Nori, Phys. Rev. A, {\bf 70}
063801 (2004); L. F. Wei, M. Zhang, H. Y. Jia, and Y. Zhao, Phys.
Rev. A, {\bf 78} 014306 (2008).
%
\bibitem{OC}
M. Zhang, H. Y. Jia, and L. F. Wei, Optics Communications, {\bf 282}
1948 (2009); C. Monroe {\it et al.}, Phys. Rev. A {\bf 55}, R2489
(1997).
%
\bibitem{experimentstate}D. M. Meekhof {\it et al.}, Phys. Rev. Lett. {\bf
76}, 1796 (1996).
%
\bibitem{experimentgate}C. Monroe {\it et al.}, Phys. Rev. Lett. {\bf
75}, 4714 (1995).
%


\end{thebibliography}
\end{document}